\documentstyle[12pt,psfig,a4,wider]{article}

\newcommand{\bm}{\bibitem}
\newcommand{\ud}{\bf}
\begin{document}
\begin{titlepage}
\renewcommand{\thefootnote}{\fnsymbol{footnote}}
%
%

\begin{center}
{\LARGE          {\sf  CENTRE ~FOR ~NUCLEAR ~PHYSICS            }}\\[1ex]
\vspace*{0.3cm}
{\Large   {\sf        UNIVERSITY~OF~SURREY                   }}\\[1ex]
\vspace*{0.2cm}
{\large          {\sf GUILDFORD, SURREY~GU2~5XH,~U.K. }}\\[1ex]
{\footnotesize     {\sf   Tel. +44 1483 300800 FAX +44 1483 259501 }}\\[1ex]
{\footnotesize     {\sf email: I.Thompson@ph.surrey.ac.uk}}\\[1ex]
\vspace*{1.0cm}

\end{center}
\vspace*{-1.5cm}
\hrule

\vspace*{0.4cm}
\centerline{\large\sf Preprint CNP-95/18}
%

%
{\sf
  
\vspace{0.6cm}

\begin{center}
{\LARGE\bf Astrophysical S-factor of the
$^7$Be($p,\gamma)^8$B reaction
from Coulomb dissociation of $^8$B\footnote{Work supported by EPSRC, UK, grant
no. GR/K53048.}}\\[1.0cm]
{\bf R. Shyam\footnote{ Permanent Address: Saha Institute of Nuclear Physics,
Calcutta, India} and I.J. Thompson}\\ 
{\it Department of Physics, University of Surrey, Guildford,\linebreak Surrey GU2 5XH, 
U.K.}\\ 
{\bf A.K. Dutt-Mazumder}\\
{\it Saha Institute of Nuclear Physics, 
Calcutta - 700 064, India}\\[1.0 cm]
\end{center}
\renewcommand{\thefootnote}{\arabic{footnote}}
\begin{abstract}
\small
The Coulomb dissociation method to obtain the astrophysical
S-factor, $S_{17}(0)$, for the $^7$Be($p,\gamma)^8$B  reaction at
solar energies is investigated by analysing the recently
measured data on the breakup reaction 
$^{208}$Pb$(^8$B,$^7$Be$~p)^{208}$Pb at 46.5 MeV/A beam energy.
Breakup cross sections corresponding to 
E1, $E2$ and $M1$ transitions are calculated with a theory of Coulomb   
excitation that includes the effects of the Coulomb recoil 
as well as relativistic retardation. The interplay of nuclear   
and Coulomb contributions to the breakup process is studied
by performing a full quantum mechanical calculation within the 
framework of the distorted-wave Born Approximation. In the kinematical
regime of the present experiment, both nuclear as well as Coulomb-nuclear 
interference processes affect the pure Coulomb breakup cross sections 
very marginally. The $E2$ cross sections are strongly dependent on
the model used to describe the structure of $^8$B.
The value of $S_{17}(0)$ is deduced  
with and without $E2$ and $M1$ contributions added to the $E1$    
cross sections and the results are discussed.  

 

\bigskip

\centerline{\large\bf Physics Letters B, in press.  }
\end{abstract}
}

\end{titlepage}
The radiative capture reaction $^7$Be$(p,\gamma )^8$B plays a
crucial role in the so called ``Solar neutrino puzzle".  The
cross sections of this reaction at solar energies 
($E_{cm}\simeq 20$ keV where $E_{cm}$ is
the center of mass relative energy of the $p + ^7Be $ system)
are directly related to the flux of the high energy neutrinos 
(emitted in the subsequent $\beta$ decay of $^8$B) to which the
$^{37}$Cl  and Kamiokande detectors are most sensitive [1]. The
direct measurement of this reaction, at such low relative
energies, is strongly hindered because of the Coulomb barrier.
Therefore, the cross sections measured at relatively higher
values of $E_{cm}$ are extrapolated to solar energies using a
theoretically derived energy dependence of the low energy cross
sections.  However, absolute values of the S-factors
obtained by various workers by following this method 
differ from one another considerably [2,3,4].

The method of Coulomb dissociation provides an alternate
indirect way of determining the cross sections of the radiative
capture reaction at low relative energies. In this procedure it
is assumed that the break-up reaction
$a+ Z \rightarrow (b + x) + Z$ proceeds entirely via
the electromagnetic interaction; the two
nuclei $a$ and $Z$ do not interact strongly.
By further assumption that the electromagnetic excitation
process is of first order, one can relate [5] directly the
measured cross-sections of this reaction to those of the
radiative capture reaction $b + x \rightarrow a + \gamma $.
Thus, the astrophysical S-factors of the radiative capture processes
can be determined from the study of break-up reactions under
these conditions.  A few reactions e.g. 
$\alpha + d \rightarrow ^6{\rm Li} + \gamma$ , 
$\alpha + t \rightarrow ^7{\rm Li} + \gamma$, and 
$^{13}{\rm N} + p \rightarrow ^{14}{\rm N} + \gamma $
have been studied both experimentally and theoretically using
this method [see e.g. Ref. [6] for a recent review].

A first attempt has been made recently by Motobayshi et al. [8],
to study the Coulomb dissociation of $^8$B into the $^7$Be$-p$ low
energy continuum in the field of $^{208}$Pb with a radioactive
$^8$B beam of 46.5 MeV/A energy [7]. Assuming a pure E1
excitation, the Monte Carlo simulation of their data predicts a
$S_{17}(0)=16.7 \pm 3.2$ eV barn, which is considerably
lower than the value of 22.4 $\pm$2.0 eV barn used by
Bahcall and Pinsonneault [8] in
their standard solar model (SSM) calculations.

In this letter we
perform a more rigorous analysis of the data of Ref. [7] by using
a theory of Coulomb excitation which simultaneously 
includes the effects of Coulomb recoil and relativistic retardation  
by solving the general classical problem of the motion of two  
relativistic charged particles [9]. Under the kinematical condition
of the present experiment, $E2$ transitions may be disproportionately
enhanced in the Coulomb dissociation process. There has been
quite some debate in the literature recently [10,11,12] about the extent of E2
contribution to the data of Ref. [7]. We would, therefore, like to
reexamine this issue carefully, as this has important consequences
for the $S_{17}(0)$ extracted from the data.   
We also include the contributions of the $M1$ transition which may be 
important for $E_{cm}$ in the vicinity of 0.633 MeV
which corresponds to a $1^{+}$ continuum resonant state in $^8$B.
Furthermore, we investigate the role of the nuclear
excitations. Although the cross sections of the pure nuclear
breakup may be small as compared to those of the Coulomb
break-up, their interference may still have some effect on the angular 
distributions. The usefulness of the Coulomb 
dissociation method in extracting the reliable astrophysical $S$ factor 
from the breakup data 
depends on this term having negligible influence on the 
calculated break-up cross sections. We report here the result
of the first full quantum mechanical calculation of the Coulomb
and nuclear excitations and their interference for this reaction,
performed within the framework of the distorted wave
Born approximation (DWBA). This is expected to highlight whether or not 
the nuclear effects alter appreciably the prediction of a pure Coulomb
breakup process.   

The double differential cross-section for the Coulomb excitation
of a projectile from its ground state to the continuum, with a
definite multipolarity of order $\pi\lambda$ is given by [5,6]  
\begin{eqnarray} 
\frac{d^2 \sigma}{d \Omega dE_{\gamma}} & = & \sum_{\pi\lambda}\frac{1}{E_\gamma}
                       \frac{dn_{\pi \lambda}}{d \Omega} \sigma_{\gamma}^{\pi
                              \lambda}(E_\gamma),
\end{eqnarray}
where
$\sigma_{\gamma}^{\pi \lambda}(E_{\gamma})$ is the cross-section for
the photodisintegration process $ \gamma + a \rightarrow b + x$,
with photon energy $E_\gamma$, and multipolarity
$\pi\,=\,$ E (electric) or M (magnetic), and 
$\lambda\,=\, 1,2...$ (order), which is related to that of
the radiative capture process $\sigma (b + x
\rightarrow a + \gamma)$ through the theorem of detailed
balance. In terms of the astrophysical S-factor, we can write 
\begin{eqnarray}
\sigma (b + x \rightarrow a + \gamma ) & = & \frac{S(E_{cm})}{E_{cm}}
                                      \exp(-2 \pi \eta(E_{cm})),
\end{eqnarray}
where $\eta\, =\,\frac{Z_b Z_x e^2}{\hbar v}$,
with $v$, $Z_b$ and $Z_x$ being the relative center of mass
velocity, and charges of the fragments $b$ and $x$ respectively.

In most cases, only one or two multipolarities dominate the 
radiative capture as well as the Coulomb dissociation cross sections.
In Eq. (1)
$n_{\pi\lambda}(E_\gamma)$ represents the number of equivalent
(virtual) photons provided by the
Coulomb field of the target to the projectile, which is  
calculated by the method discussed in Ref. [9].   
$S(E_{cm})$ can be 
directly determined from the measured Coulomb dissociation
cross-sections using Eqs. 1 and 2.

The quantal treatment of the Coulomb and nuclear excitations is
well known [13]. The expression for the differential cross sections
corresponding to a transition of multipolarity $\lambda$ can be
schematically represented as
\begin{eqnarray}
\frac{d\sigma}{d\Omega} & \propto & \sum_{\lambda=-\mu}^{\lambda=+\mu}
                       |\int_{0}^{\infty}dr R_{L_i}(k_ir)(F_{C}^{\lambda} 
                       + F_{N}^{\lambda}) R_{L_f}(k_fr)|^2
\end{eqnarray}
In Eq. (3), indices $i$ and $f$ refer to the incoming and outgoing channels.
This equation also involves summation over the partial waves $L_i$ and $L_f$.
$F_{C}^\lambda$ and $F_N^\lambda$ denote the form factors for the
Coulomb and nuclear excitations, respectively. The expression
for $F_{C}^{\lambda}$ which is determined entirely by the corresponding
$B(E\lambda)$ value, is given in Refs. [13,14]. For $F_{N}^\lambda$, we take the 
usual collective model expression with the value of the
``nuclear deformation parameter'' being the same as that of
the Coulomb one. 
$R_{L_i}$ and $R_{L_f}$ define
the wave functions of the relative motion in the incoming and outgoing
channels, respectively, which are obtained by solving the Schr\"odinger
equation with appropriate optical potentials. In our calculations
we have used the potentials given in table 3 
of Ref.[15] (set D).
The same set of potentials were used for the
incoming and outgoing channels. For a Glauber-model calculation of the
Coulomb and nuclear breakup we refer to Ref. [16].

\begin{figure}[htb]
\centerline{\psfig{figure=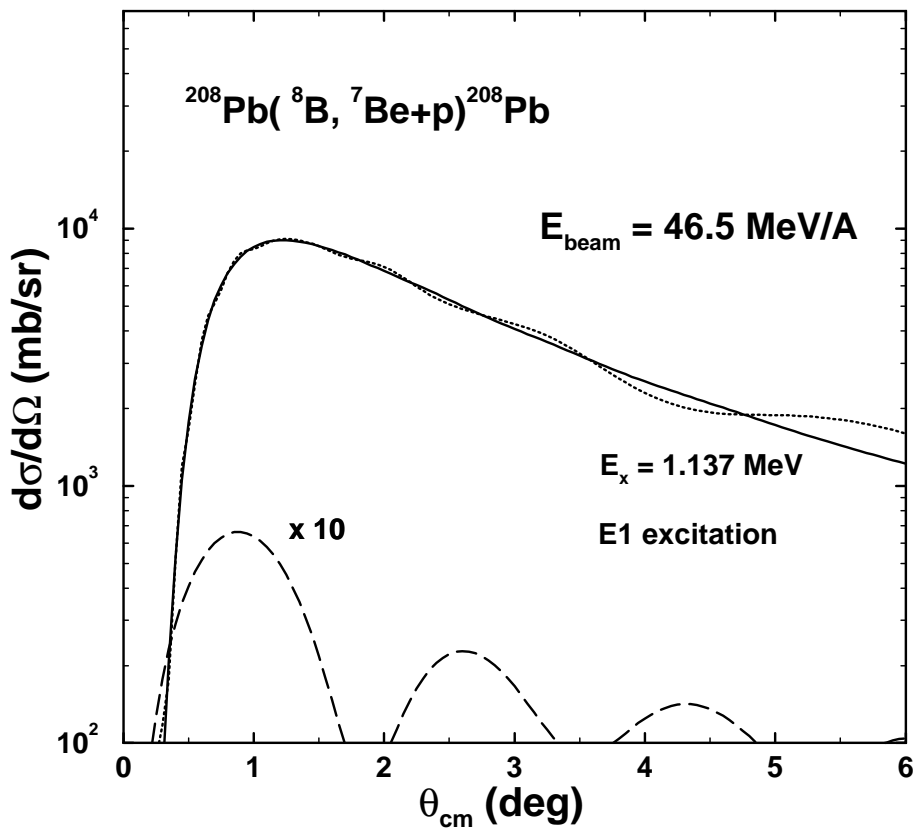,width=10cm}}
\vspace{1pt}
\caption[]{
Angular distribution for 
$^8$B$ + ^{208}$Pb$ \rightarrow ^8$B$^* (^7$Be$+p) +^{208}$Pb reactions
for an $E1$ transition corresponding to an
excitation energy of 1.137 MeV. The dashed (dotted)
curve represent the result of pure nuclear (Coulomb) excitation 
calculation. Their coherent sum is depicted by solid curve. The 
scattering angle corresponds to the  center-of-mass
angle ($\theta_{cm}$) of outgoing $^8B^*$.
}
\label{fig1}
\end{figure}

In Fig. 1, we show the results of our full quantal calculations for the
E1 breakup of $^8$B on $^{208}$Pb target at the beam energy
of 46.5 MeV/A for $E_{cm}$ of 1.0 MeV. We have chosen a higher value
of $E_{cm}$ where nuclear effects may be larger; the calculation 
reported in Ref. [16] is at a very low value of $E_{cm}$ (100 keV). 
The solid line shows the result obtained with the coherent sum of
nuclear and Coulomb excitations as well as their interference. The
dashed (dotted) line depict the cross sections for pure nuclear
(Coulomb) excitation. We have used 5000 partial waves ($L_i$ and $L_f$)
in the calculation of the Coulomb excitation
cross section. We note that the
pure Coulomb excitation angular distribution   
is a smooth curve approaching zero as $\theta \rightarrow 0$ (the
so called adiabaticity limit). This reflects the semiclassical nature
of the process, thus the pure Coulomb excitation process should
be amenable to the semiclassical methods [17]. 
The peak in this distribution corresponds to an angle $\theta_{min}$ 
($\simeq \frac{2aE_\gamma}{\hbar v}$, where $a$ is half the distance
of closest approach in a head on collision) 
below which the adiabaticity condition sets in. On the other
hand, the pure nuclear cross sections show the typical
diffraction pattern, and are at least two orders of magnitude
smaller than the pure Coulomb one. The important point to note is that
pure Coulomb cross sections are very similar to full calculations. 
Similar results are obtained also for the values of $E_{cm}$
of 0.8 MeV and 0.6 MeV. 
Thus, nuclear effects modify the total amplitudes
very marginally in the entire kinematical regime of the data of Ref.
[7] (except for the region very close to $\theta = 0$). Thus the Coulomb 
dissociation method can be used to extract reliable 
the astrophysical S factor for the $^7Be(p,\gamma)^8B$ reaction from
this data.     

\begin{figure}[p]
\centerline{\psfig{figure=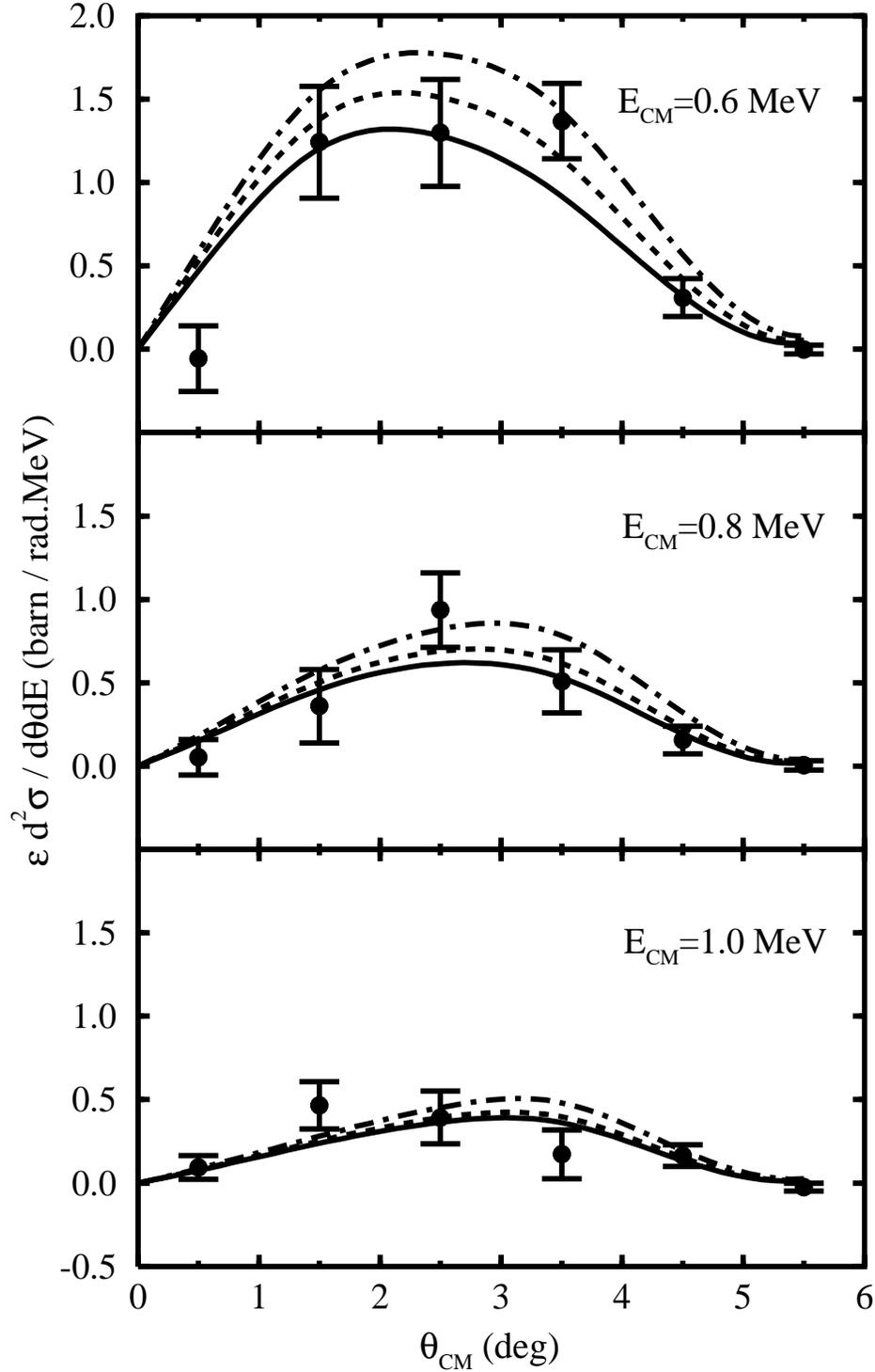,height=19cm}}
\vspace{1pt}
\caption[]{
Comparison of experimental and theoretical 
Coulomb  dissociation yields (cross section $\times$ detector
efficiency) as a function of $\theta_{cm}$ for the $E_{cm}$ 
values of 0.6 MeV, 0.8 MeV and 1.0 MeV.
Solid lines show the calculated pure $E1$ Coulomb dissociation
cross sections obtained with best fit 
values of S factors as discussed in the text. The dashed and dashed dotted
curves represent the sum of $E1$, $E2$ and $M1$ contributions with 
latter two components calculated with capture cross sections 
given in the models of TB [18]
and KPK [19] respectively. The experimental data is taken from
Ref. [7].
}
\label{fig2}
\end{figure}
   
In Fig. 2, we show the comparison of our calculated Coulomb dissociation 
double differential cross sections with the corresponding data of Ref. [7]  
as a function of the scattering angle $\theta_{cm}$ of the
excited $^8$B (center of mass of the $^7$Be$ + p $ system) for
three values of the $E_{cm}$. The calculated $E1$, $E2$ and $M1$ cross sections 
are folded with an efficiency matrix provided to us by  Naohito
Iwasa (one of the authors of Ref. [7]). This matrix accounts
for the efficiency and geometry of the detectors
(including energy and angular spread)\footnote   
{In Ref. [7], the detector response was
taken into account by putting the theoretical cross sections 
as input to a Monte Carlo simulation program.}.      
The solid lines in Fig. 2 show the calculated $E1$ cross sections  
obtained with S-factors $(S_{17})$ that provide
best fit to the data (determined by $\chi ^2 $ minimization procedure).
These are $(17.58 \pm 2.26)$ eV barn, $(14.07 \pm 2.67)$
eV barn  and $(15.59 \pm 3.49)$ eV barn at $E_{cm}$= 0.6 MeV,
0.8 MeV and 1.0 MeV respectively. 

The contributions of the $E2$ and $M1$ excitations
are calculated by using the radiative capture cross sections 
$\sigma(p+^7$Be$\rightarrow ^8$B$ +\gamma)$, given by the models of
Typel and Baur (TB) [18] and Kim, Park and Kim (KPK) [19].
We have used as input the corresponding S factors averaged 
over energy bins of experimental uncertainty in the
relative energy of the fragments. In Fig. 2, the dashed (dashed
dotted) line shows the $E1$ (with best fit
$S_{17}$) + $E2$ + $M1$ cross sections, with $E2$ and $M1$ components 
calculated  with TB (KPK) capture cross sections .
It should be noted that the contribution of $M1$ 
component is substantial for $E_{cm}\,=\, 0.6$,
while at 0.8 MeV and 1.0 MeV it is relatively 
quite small. This  multipolarity was not included in the analysis
presented in Ref. [10] and also in the calculations
of the angular distributions shown in Ref. [20]. The $E2$ capture 
cross sections of KPK and TB are 
quite different from each other, while those of $M1$ 
multipolarity are (within 10-15$\%$) model independent.
If the KPK model is correct then $E2$ contributions to the data
of Ref. [7] is not negligible.
On the other hand, with the TB model the calculated ($E1$+$E2$+$M1$) 
cross sections are within the experimental uncertainties of the data.

The best fit ``$E1$ only'' $S_{17}$ factors, as described above, 
give a $S_{17}(0) \,= \,(15.5 \pm 2.80)$ eV barn,  
if we use the direct extrapolation procedure adopted in 
Ref. [7] together with the TB capture model. It should be noted
that in this model the E1 capture from both $s$ and $d$ waves
to the $p$-wave ground state  is included. In the model of
Tombrello [21], which is the basis of extrapolation in Ref.[7],
the $d$ wave contributions are ignored. This can lead to
a 10-15$\%$ difference in the extrapolated S factor.
To see the effect of the $E2$ and $M1$ components on the extracted
astrophysical S factor we refitted the data of Ref. [7] with
($E1$(best fit) + $E2$ + $M1$) multiplied by a parameter $\xi (E_{cm})$, 
which is determined by $\chi^2$ minimization
procedure. The best fit values of the ``correction factor'' $\xi$
are $(0.70 \pm 0.17)$, $(0.73 \pm 0.10)$ and $(0.75 \pm 0.15)$ with 
the KPK model and $(0.85 \pm 0.12)$, $(0.88 \pm 0.16)$
and $(0.91 \pm 0.20)$ with the TB model, at the relative energies
of 0.6 MeV, 0.8 MeV and 1.0 MeV respectively.
The values of the parameter $\xi(E_{cm})$ translated into $E1$ 
S factors give a corrected $S_{17}(0)$ of $11.20 \pm 2.02$ 
($14.0 \pm 2.45$) for KPK (TB) model. Thus, with
the KPK model, the $E2$ and $M1$ contributions reduce the ``E1 only S-factor''
by more than $25 \%$, while with TB model
this reduction is limited to only about $10-12 \%$.

\begin{figure}[tbhp]
\centerline{\psfig{figure=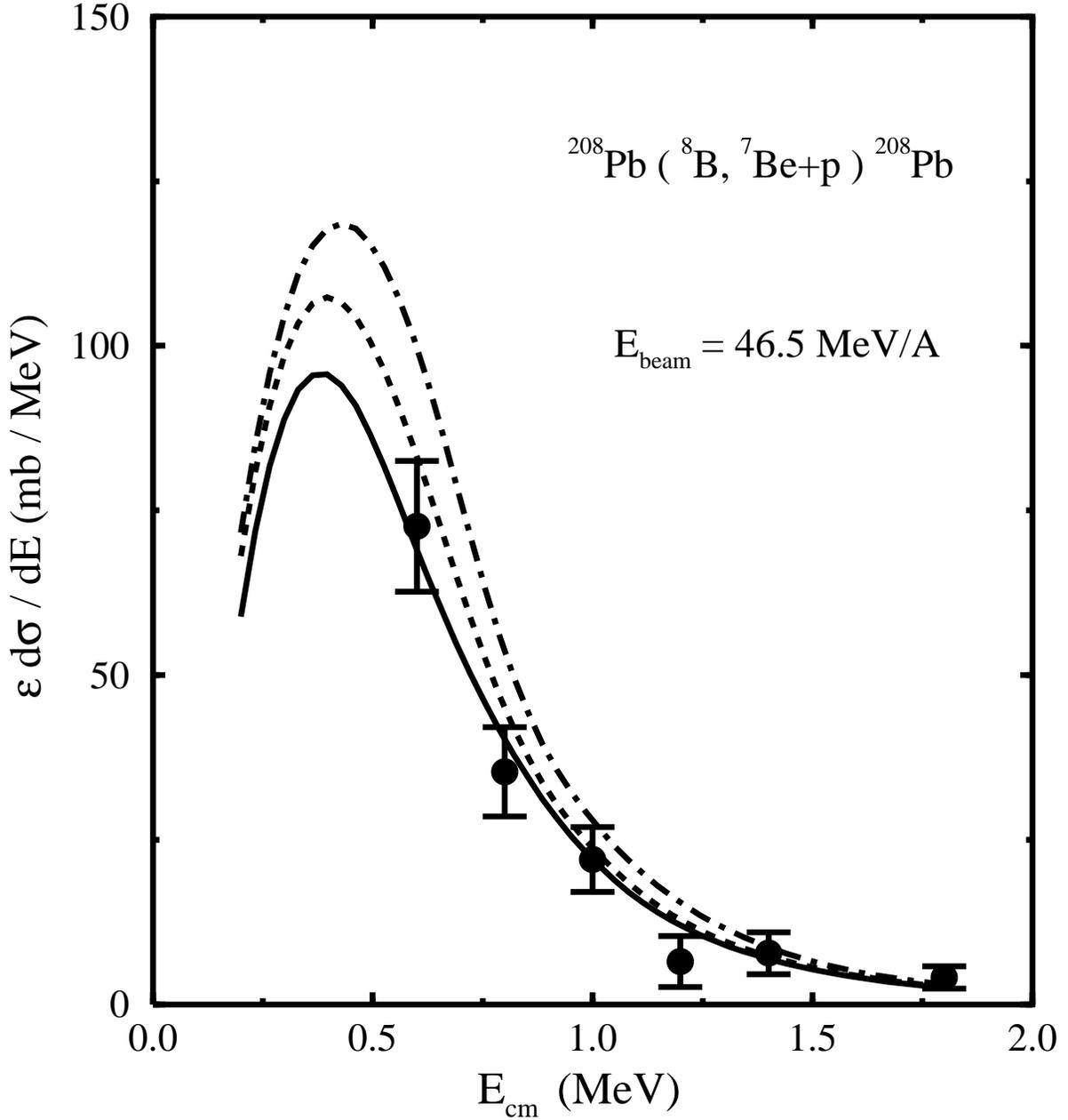,width=14cm}}
\vspace{1pt}
\caption[]{
Comparison of the experimental and calculated angle
integrated Coulomb dissociation yields as a function of fragment
relative energy. The solid curve represents pure $E1$ cross sections 
calculated with a constant S factor of 15.5 eV barn. Dashed and dashed dotted 
lines show the sum of the $E1$, $E2$ and $M1$ components with latter two 
being calculated with capture cross sections predicted by TB
and KPK respectively. The data is from Ref. [7].
}
\label{fig3}
\end{figure}

In fig. 3, the calculation for the angle integrated cross sections
is compared with the data. The solid line shows the results for the 
$E1$ breakup calculated with a constant astrophysical S factor of
15.5 eV barn. The dashed (dashed dotted) line shows the  
sum of the $E1$, $E2$ and $M1$ contibutions with latter two components 
calculated with the capture cross sections of TB (KPK).
It may be noted that unlike the case in Ref. [20],
we have used efficiencies corresponding to $E2$ and $M1$ multipolarities
to fold the calculated cross sections of these transitions.
We observe that with the KPK model the cross sections 
up to $E_{cm} \simeq 1.2$ MeV are modified by the 
$E2$ and $M1$ components, while with the TB model only the
point at 600 keV is appreciably modified. 

Thus we see that $E2$ corrections to the data of Ref. 7 is 
strongly model dependent and it is difficult to
draw any definite conclusion about their contributions. 
In the past the KPK model  has been criticised 
by Barker [22] for their incorrect  treatment of the resonant
contribution due to $M1$ and $E2$ transitions. On the other hand
Typel and Baur have neglected the capture from $f$-wave relative
initial states, which could lead to a smaller $E2$ capture cross section.
The calculations of several other authors also differ in their 
predictions of the $E2$ cross section [23]. It is therefore 
important to develop a reliable model to calculate the $E2$ 
capture cross sections for the $^7$Be$(p,\gamma)^8$B reactions.

\begin{figure}[p]
\centerline{\psfig{figure=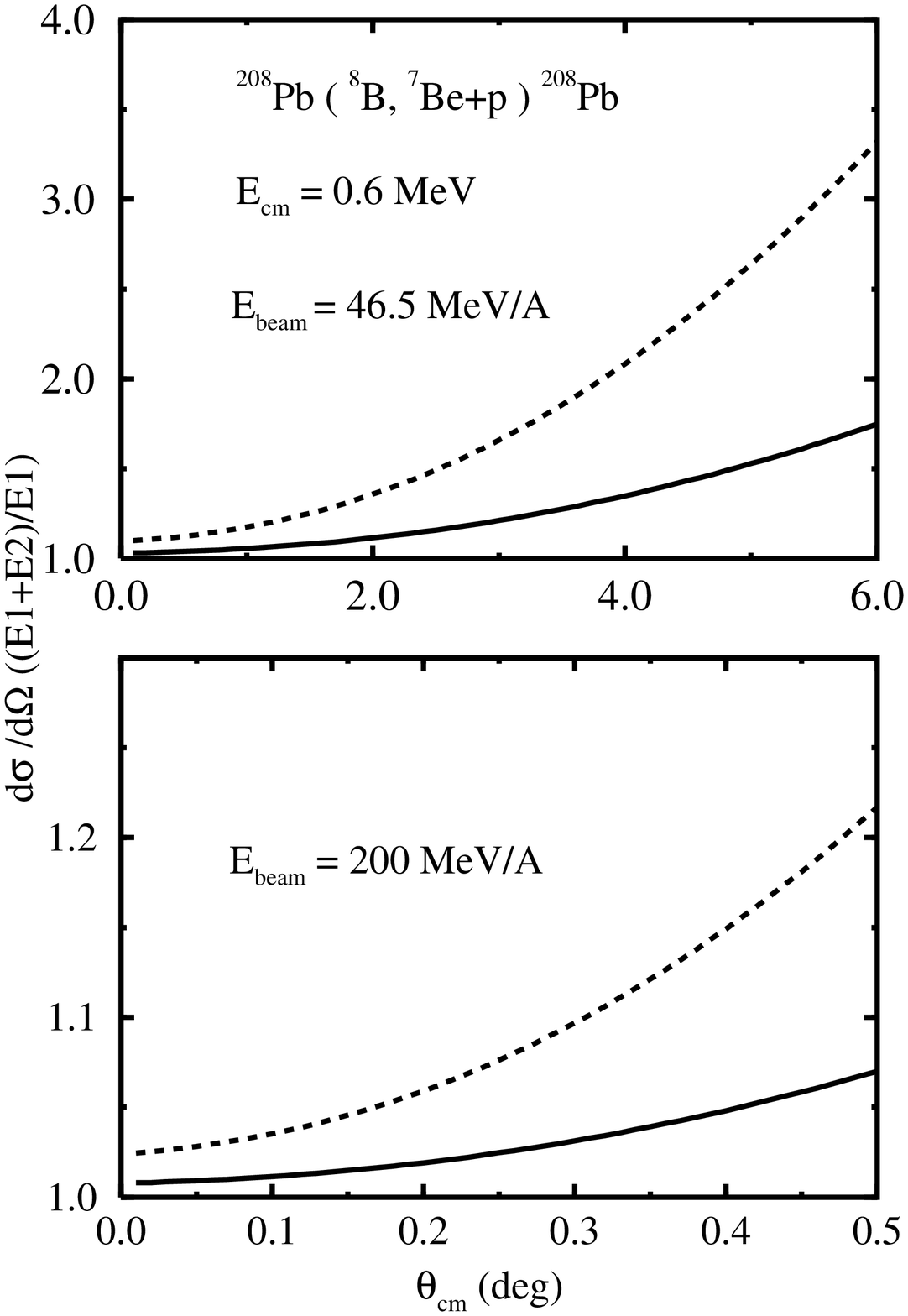,height=15cm}}
\vspace{1pt}
\caption[]{
The ratio of differential cross sections for $E1$ + $E2$ transitions
to pure $E1$ transition (calculated with a S factor of 15.5 eV barn)
as a function of $\theta_{cm}$ for the reactions
of Fig. 1 at the beam energies of 46.5 MeV/A (upper part) and 200 MeV/A (lower part).
The solid
(dashed) line corresponds to the case where $E2$ component has been calculated 
with the capture cross sections of TB (KPK).}
\label{fig4}
\end{figure}

Nevertheless, measurements performed at higher beam energies 
are likely to provide a kinematical regime in which Coulomb
dissociation cross section are less dependent on the
nuclear structure model, as $E2$ components in this regime are
appreciably weaker than their $E1$ counterpart. In Fig. 4
we show the ratio $d\sigma/d\Omega ((E1+E2)/E1)$ at the beam 
energies of 46.5 MeV/A and 200 MeV/A as a function of 
$\theta_{cm}$, with $E2$ cross sections
calculated within the KPK(dotted lines) and TB 
(solid lines) models for the $E_{cm}$ value of  
0.6 MeV. The $E1$ angular distributions ($d\sigma/d\Omega$)
decrease as $\theta_{cm}$ 
increases beyond $\theta_{min}$ while those of
E2 multipolarity remains flat (in the angular region 
considerd in this figure). Thus this ratio increases with angle;
the rate of the increase being determined by the magnitude
of the $E2$ component. We notice that the value of this ratio is around 3.0
(1.2) at the beam energy of 46.5 MeV/A (200 MeV/A) with $E2$ cross sections
calculated in the KPK model. This implies that at 
the beam energies around 200 MeV and at very forward angles 
the $E2$ contributions calculated even in the KPK model 
will introduce relatively small 
modifications to pure $E1$ cross sections (they are small in the TB model
anyway). Therefore, higher
beam energies and very forward angles are expected to be better
suited for extracting the astrophysical S-factors independent 
of the uncertainties in the nuclear strucure model of $^8$B.
Furthermore, the higher order effects (eg. the so called
post-acceleration) which can distort the the relative energy
spectrum of $^7Be + p$ system are expected to be insignificant
at these energies [20].

In summary, we analysed the recently measured data [7] of the
breakup of $^8$B on $^{208}$Pb target at the beam
energy of 46.5 MeV/A by the Coulomb dissociation method 
in order to extract the astrophysical S factors for the 
radiative capture reaction $^7Be(p,\gamma)^8B$. We used the
first order perturbation
theory of Coulomb excitation which included both the effects of the
relativistic retardation as well as Coulomb recoil simultaneously.
We considered the breakup due to $E1$, $E2$ and $M1$ transitions.
Full quantum mechanical calculations within the framework of the DWBA, 
were performed for the pure Nuclear breakup, pure Coulomb breakup 
and their interference and it is shown that nuclear break-up
contributions affect the total break-up 
amplitudes only very marginally in the kinematical regime of 
the data of Ref. [7]. Thus the assumptions of the Coulomb dissociation
method are well fulfilled for this case.

The $S_{17}(0)$ value deduced in
our analysis considering only the $E1$ cross sections is in agreement
with the lower values reported from the extrapolation of the 
direct capture cross sections, and with that extracted from a recent
indirect theoretical method [24]. This is significantly smaller
than the value used by Bahcall and 
Pinsenault [8] and Turck-Chieze et al. [1] 
in their SSM calculations. The $E2$ break-up cross
sections can reduce the ``E1 only S factors'', bringing the 
$S_{17}(0)$ further down. However, the 
magnitude of the $E2$ contributions to the data of Ref. [7]
depend significantly on the nuclear structure model used to
calculate the corresponding capture cross sections and at this stage 
it is not possible to make a definite prediction.
Experiments performed at higher beam energies are expected to be
relatively less affected by the $E2$ component, and therefore
they provide a better kinematical regime
where the extracted S factors
are likely to be less model dependent. 

One of the authors (RS) would like to thank 
Prof. R.C. Johnson and Dr. J.A. Tostevin for
their kind hospitality at the 
University of Surrey and for several useful discussions 
on the subject matter of this paper. Thanks are also
due to Drs. Naohito Iwasa and T. Motobayashi
for providing us the detector efficiencies and the experimental 
cross sections of their experiment.


\end{document}